\documentclass[twocolumn,twocolappendix]{aastex631}

\usepackage{amsmath}
\usepackage{bm}

\usepackage{soul}

\shorttitle{Cosmological coupling of black holes: first observation and implications}
\shortauthors{Farrah, Croker, et al.}

\newcommand{\uhm}{Department of Physics and Astronomy, University of Hawai`i at M\=anoa, 2505 Correa Rd., Honolulu, HI, 96822, USA}

\begin{document}

\title{Observational evidence for cosmological coupling of black holes and its implications for an astrophysical source of dark energy}

\correspondingauthor{Duncan~Farrah}
\email{dfarrah@hawaii.edu}

\author[0000-0003-1748-2010]{Duncan~Farrah}
\affiliation{Institute for Astronomy, University of Hawai`i,  2680 Woodlawn Dr., Honolulu, HI, 96822, USA}
\affiliation{\uhm}

\author[0000-0002-6917-0214]{Kevin~S.~Croker}
\affiliation{\uhm}

\author[0000-0003-1704-0781]{Gregory~Tarl\'e}
\affiliation{Department of Physics, University of Michigan, 450 Church St., Ann Arbor, MI, 48109, USA}

\author[0000-0002-2601-1870]{Valerio Faraoni}
\affiliation{Department of Physics \& Astronomy, Bishop’s University, Sherbrooke Quebec Canada J1M 1Z7}

\author[0000-0003-0624-3276]{Sara~Petty}
\affiliation{NorthWest Research Associates, 3380 Mitchell Ln., Boulder, CO 80301, USA}
\affiliation{Convent \& Stuart Hall Schools of the Sacred Heart, 2222 Broadway, San Francisco, CA 94115, USA}

\author[0000-0002-9149-2973]{Jose Afonso}
\affiliation{Departamento de F\'{i}sica, Faculdade de Ci\^{e}ncias, Universidade de Lisboa, Portugal}
\affiliation{Instituto de Astrof\'{i}sica e Ci\^{e}ncias do Espa\c co, Universidade de Lisboa, Portugal}

\author[0000-0002-3573-339X]{Nicolas Fernandez}
\affiliation{NHETC, Department of Physics and Astronomy, Rutgers University, Piscataway, NJ 08854, USA}

\author[0000-0001-8818-8922]{Kurtis~A.~Nishimura}
\affiliation{\uhm}

\author[0000-0001-6139-649X]{Chris Pearson}
\affiliation{RAL Space, STFC Rutherford Appleton Laboratory, Didcot, Oxfordshire OX11 0QX, UK}
\affiliation{Oxford Astrophysics, University of Oxford, Keble Rd, Oxford OX1 3RH, UK}
\affiliation{The Open University, Milton Keynes MK7 6AA, UK}

\author[0000-0002-6736-9158]{Lingyu Wang}
\affiliation{Kapteyn Astronomical Institute, University of Groningen, Postbus 800, 9700 AV Groningen, the Netherlands}
\affiliation{SRON Netherlands Institute for Space Research, Landleven 12, 9747 AD, Groningen, the Netherlands}

\author[0000-0002-0147-0835]{Michael~Zevin}
\affiliation{Enrico Fermi Institute, The University of Chicago, 933 East 56th Street, Chicago, IL 60637, USA}
\affiliation{Kavli Institute for Cosmological Physics, The University of Chicago, 5640 South Ellis Avenue, Chicago, IL 60637, USA}

\author[0000-0002-9548-5033]{David L Clements}
\affiliation{Imperial College London, Blackett Laboratory, Prince Consort Road, London, SW7 2AZ, UK}

\author[0000-0002-2612-4840]{Andreas Efstathiou}
\affiliation{School of Sciences, European University Cyprus, Diogenes Street, Engomi, 1516 Nicosia, Cyprus}

\author[0000-0003-0917-9636]{Evanthia Hatziminaoglou}
\affiliation{ESO, Karl-Schwarzschild-Str 2, D-85748 Garching bei München, Germany}

\author[0000-0002-3032-1783]{Mark Lacy}
\affiliation{National Radio Astronomy Observatory, Charlottesville, VA, USA}

\author[0000-0003-0639-025X]{Conor McPartland}
\affiliation{Cosmic Dawn Center (DAWN) Denmark}
\affiliation{Niels Bohr Institute, University of Copenhagen, Jagtvej 128, DK 2200 Copenhagen, Denmark}

\author[0000-0002-5206-5880]{Lura K Pitchford}
\affiliation{George \& Cynthia Woods-Mitchell Institute for Fundamental Physics and Astronomy, Texas A\&M University, TX, USA}
\affiliation{Department of Physics and Astronomy, Texas A\&M University, College Station, TX, USA} 

\author[0000-0002-4702-0864]{Nobuyuki Sakai}
\affiliation{Graduate School of Sciences and Technology for Innovation, Yamaguchi University, Yamaguchi 753-8512, Japan}

\author[0000-0002-1888-8744]{Joel Weiner}
\affiliation{Department of Mathematics, University of Hawai`i at M\=anoa, 2565 McCarthy Mall, Honolulu, HI 96822}

\begin{abstract}
Observations have found black holes spanning ten orders of magnitude in mass across most of cosmic history.
The Kerr black hole solution is however provisional as its behavior at infinity is incompatible with an expanding universe.  
Black hole models with realistic behavior at infinity predict that the gravitating mass of a black hole can increase with the expansion of the universe independently of accretion or mergers, in a manner that depends on the black hole's interior solution.
We test this prediction by considering the growth of supermassive black holes in elliptical galaxies over $0<z\lesssim2.5$.
We find evidence for cosmologically coupled mass growth among these black holes, with zero cosmological coupling excluded at 99.98\% confidence.
The redshift dependence of the mass growth implies that, at $z\lesssim7$, black holes contribute an effectively constant cosmological energy density to Friedmann's equations.
The continuity equation then requires that black holes contribute cosmologically as vacuum energy.
We further show that black hole production from the cosmic star formation history gives the value of $\Omega_{\Lambda}$ measured by Planck while being consistent with constraints from massive compact halo objects.
We thus propose that stellar remnant black holes are the astrophysical origin of dark energy, explaining the onset of accelerating expansion at $z \sim 0.7$. 
\end{abstract}


\section{Introduction}\label{sec:intro}
Astrophysical black holes (BHs), with masses spanning a few to several billion solar masses, are found in systems ranging from stellar binaries to supermassive BHs in the centers of galaxies.
Observations of gravitational waves from binary BH mergers \citep{abbott2019gwtc,abbott2021tests} and of supermassive BHs by the \citet{ehtm8719} and \citet{akiyama2022first}, have shown excellent consistency with the \citet{kerr1963gravitational} solution on timescales from milliseconds to hours, and spatial scales of up to milliparsecs.
BHs are thus established as a universal phenomenon, across at least ten orders of magnitude in mass.
 
Existing models for astrophysical BHs are necessarily provisional. They feature singularities, horizons, and unrealistic boundary conditions%
\footnote{
  Boundary conditions include both behavior at junctions and asymptotic behavior at infinity, if applicable.
}
\citep[e.g.][]{wiltshire2009kerr}.
Though singularities and horizons are of theoretical interest \citep[e.g.][]{harlow2016jerusalem}, the Kerr solution reduces to flat spacetime at spatial infinity.
This is incompatible with our universe, which is in concordance with a perturbed Robertson–Walker (RW) cosmology to sub-percent precision \citep[e.g.][]{aghanim2020planck, dodelsoncosmology2020}.
Thus, regardless of singularities and horizons, Kerr is only appropriate for intervals of time short compared to the reciprocal expansion rate of the universe, and can only be consistently interpreted as an approximation to some more general solution.

\begin{figure*}
  \includegraphics[width=0.95\linewidth]{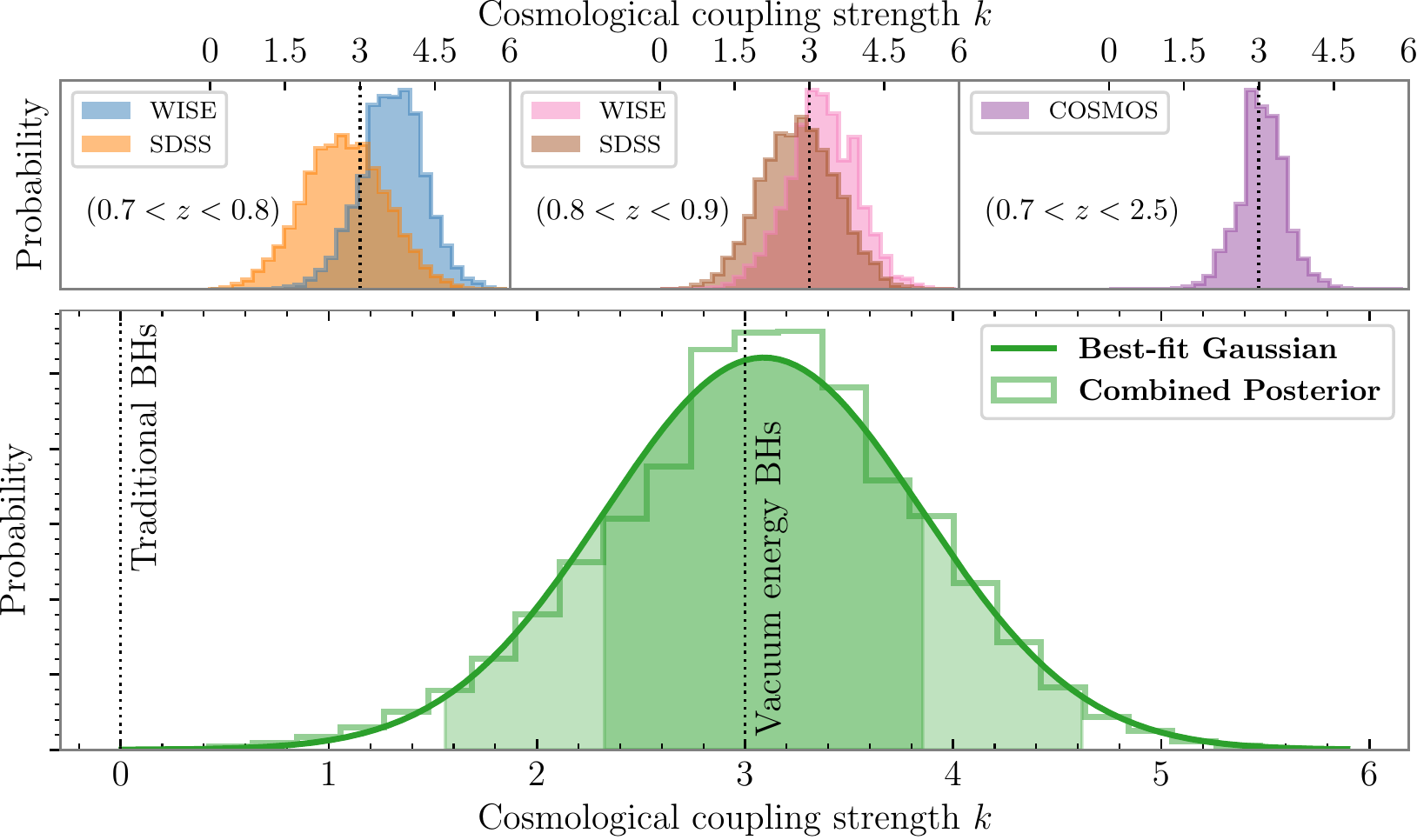}
  \caption{\label{fig:couplingpost} (Top) Posterior distributions of cosmological coupling strength $k$, inferred by comparing SMBHs in local elliptical galaxies to those in five samples of elliptical galaxies at $z>0.7$.
    (Bottom) Combining these posterior samples with equal weighting gives a distribution with $k=3.11^{+1.19}_{-1.33}$ at $90\%$ confidence.
    If fit to a Gaussian, the fit has a mean of $k=3.09$ with a standard deviation of $0.76$ (shading).
    Vertical lines indicate: $k=0$ coupling, as expected for traditional BHs like Kerr and the decoupled solution by \citet{nolan1993sources}; and $k=3$ coupling, as predicted for vacuum energy interior BHs.
    The measurement disfavors zero coupling at $99.98\%$ confidence and is consistent with BHs possessing vacuum energy interiors, as first suggested by \citet{gliner1966algebraic}.}
\end{figure*}%

Efforts to construct a BH model in General Relativity (GR) with realistic RW boundary conditions have been ongoing for nearly a century, but have met with limited success.
Early work by \citet{mcvittie1933mass} generalized the Schwarzschild solution to arbitrary RW spacetimes.
\citet{nolan1993sources} constructed a non-singular interior for this solution, and progress has been made in understanding its horizon/causal structure \citep[e.g.][]{kaloper2010mcvittie, lake2011more, faraoni2012bizarre, da2013expansion}.
\citet{farja07} constructed solutions featuring dynamical phenomena such as horizons that comove with the universe's expansion, evolution of interior energy densities and pressures, and time-varying mass.
These solutions are significant, because they show how heuristic application of Birkhoff's theorem in cosmological settings can fail in the presence of strong gravity%
\footnote{
  It is not sufficient that densities and pressures exterior to some region justify a Newtonian treatment.
  Densities and pressures interior to that region must also remain non-relativistic \citep{weinberg2008cosmology}.
}
\citep[c.f.][]{lemaitre1931, einstein1945influence, callandickepeebles1965wrong, peebles1993principles}.
Time-varying mass in particular has been studied by \citet{guariento2012realistic} and \citet{maciel2015timedep}, but its interpretation remains largely unexplored.
All of these solutions, however, are incompatible with Kerr on short time-scales because they do not spin.
A BH solution that satisfies observational constraints at small and large scales simultaneously has yet to be found.

Progress on these problems in GR has become possible with advances that resolve a long-standing ambiguity in Friedmann's equations \citep[e.g.][]{ellis1987fitting}.
In RW cosmology, the metric is position-independent and has no preferred directions in space.
In order for Einstein's equations to be consistent, the stress-enegy must share these properties.
Einstein's equations, however, give no prescription for converting the actual, position-dependent, distribution of stress-energy observed at late-times into a position-independent source.
\citet{cro19} resolved this averaging ambiguity, showing how the Einstein-Hilbert action gives the \emph{necessary} relation between the actual distribution of stress-energy and the source for the RW model. 

A consequence of this result is that relativistic material, located anywhere, can become cosmologically coupled to the expansion rate.
This has implications for singularity-free BH models, such as those with vacuum energy interiors \citep[e.g.][]{gliner1966algebraic, dymnikova1992vacuum, chaplinelaughlin2002, mazur2004gravitational, lobo2006stable, mazur2015surface, dymnikova2016regular, posada2017slowly, posada2019radial, beltracchi2019formation}.
The stress-energy within BHs like these, and therefore their gravitating mass, can vary in time with the expansion rate.
The effect is analogous to cosmological photon redshift, but generalized to timelike trajectories. 

The presence or absence of cosmologically coupled mass in BHs strongly constrains observationally viable BH solutions.
In general, the way in which a BH's mass $M$ changes in time depends on the BH model.
\citet{cro21} give a parameterization of the effect in terms of the RW scale factor $a$,
\begin{align}
  M(a) = M(a_i)\left(\frac{a}{a_i}\right)^{k} \qquad a \geqslant a_i, \label{eqn:coupling}
\end{align}
where $a_i$ is the scale factor at which the object becomes cosmologically coupled and $k \geqslant 0$ is the cosmological coupling strength.%
\footnote{This model agrees with the GR prediction for a population of identical objects, with homogeneous and non-singular interiors.
    For such objects, $k=-3P/\rho$, where $P$ and $\rho$ are the object's interior pressure and energy density, respectively.
}
The \cite{nolan1993sources} solution can be regarded as cosmologically coupled with $k=0$ because its stress-energy evolves such that the mass remains fixed throughout the cosmological expansion.
Vacuum energy interior solutions with cosmological boundaries have been predicted to produce $k \sim 3$, which is the maximum value for causal material with positive energy density \citep{cro19}.

Cosmologically coupled mass change allows for experimental distinction between singular and non-singular BHs, complementing constraints from short time-scale data \citep[e.g.][]{sakai2014gravastar, uchikata2016tidal, yunes2016theoretical,cardoso2016gravitational,cardoso2017tests,chirenti2018black,konoplya2019stable,maggio2020does}.
Observing cosmologically coupled mass change, however, is challenging.
Between an initial scale factor $a_i$ and a final one $a_f$, mass evolution only becomes apparent when $(a_f/a_i)^k \gg 1$.
For example, an observational test of cosmological mass change at $z\lesssim3$ requires a population of BHs whose masses can be tracked across at least a Gyr, and in which accretion or merging can be independently estimated.

In this paper, we perform a direct test of BH mass growth due to cosmological coupling.
A recent study by \citet{farrah22b} compares the BH masses $M_{BH}$ and host galaxy stellar masses $M_*$ of `red-sequence' elliptical galaxies over $6-9$\,Gyr, from the current epoch back to $z\sim2.7$.
The study finds that the BHs increase in mass over this time period by a factor of $8-20\times$ relative to the stellar mass.
The growth factor depends on redshift, with a higher factor at higher redshifts.
Because SMBH growth via accretion is expected to be insignificant in red-sequence ellipticals, and because galaxy-galaxy mergers should not on average increase SMBH mass relative to stellar mass, this preferential increase in SMBH mass is challenging to explain via standard galaxy assembly pathways \citep[][\S5]{farrah22b}.
We here determine if this mass increase is consistent with cosmological coupling, and, if so, the constraints on the coupling strength $k$.

\section{Methods}
\begin{figure*}
  \includegraphics[width=\linewidth]{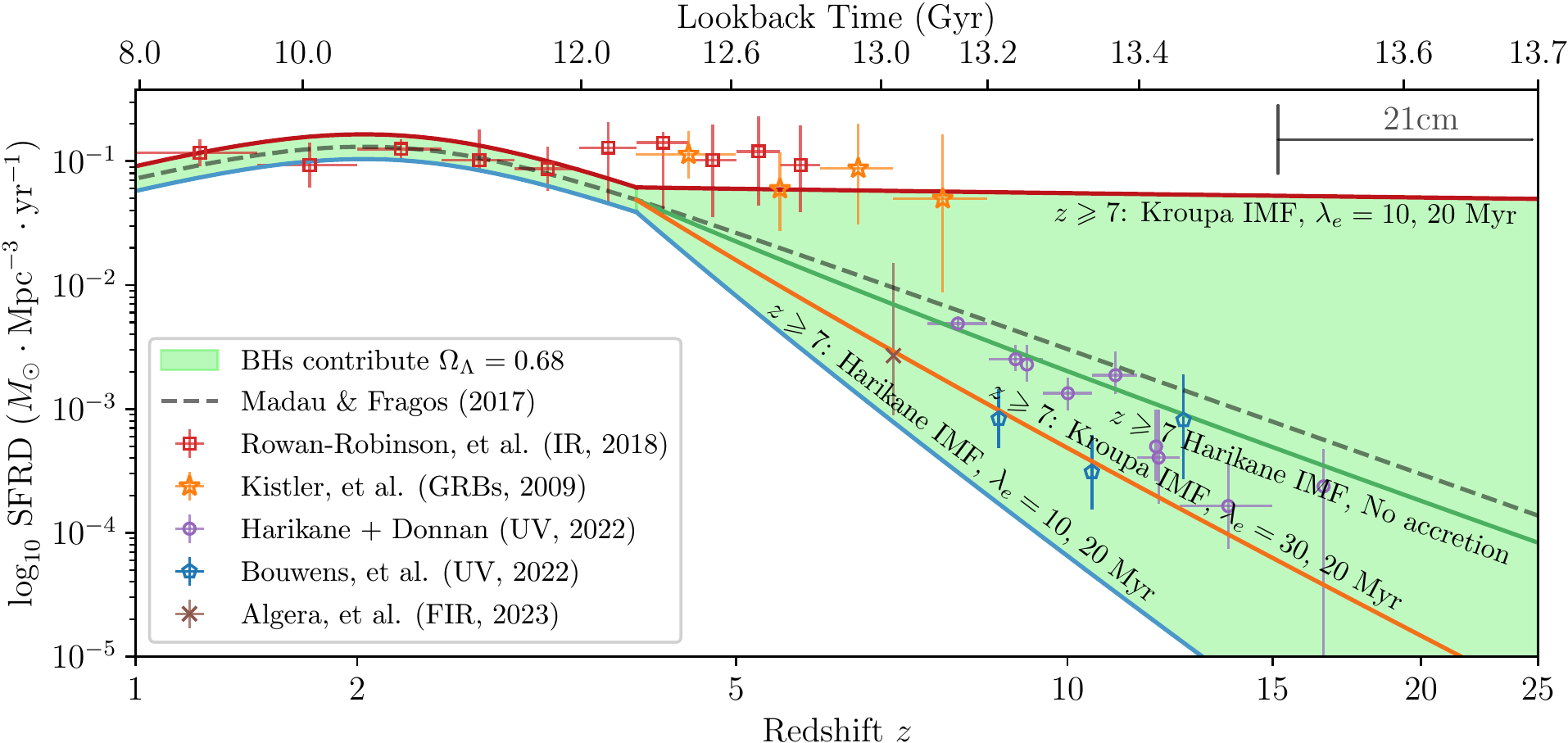}
  \caption{\label{fig:couplingsfr}
    Cosmic star formation rate densities (SFRDs) capable of producing the necessary $k=3$ BH density to give $\Omega_\Lambda=0.68$ (green, solid). 	The details of the model are given in Appendix~\ref{sec:sfrd}.
    The upper bound of the viable region adopts a \citet{kroupa03} IMF at all redshifts with the least amount of remnant accretion required to produce $\Omega_\Lambda$ with a decreasing power-law SFRD model (red, solid).
    The lower bound adopts the top-heavy IMF of \citet{harikanedropouts2022} at $z > 7$ (blue, solid).
    Two middle lines show the impact of a top-heavy IMF at $z > 7$, but no remnant accretion (green, solid); and higher accretion, but with a Kroupa IMF (orange, solid).
    Existing measurements of the SFRD via IR \citep[][red, squares]{mrr18}, $\gamma$-ray bursts \citep[][orange, stars]{kist09}, FIR \citep[][brown, xs]{almarebels2023}, and rest-frame UV via JWST \citep[][purple, dots]{donnandropouts2022, harikanedropouts2022}, \citep[][blue, dots]{bouwens2022} are over-plotted.
    The UV points can vary $\sim -1\,$dex depending upon IMF assumptions and UV luminosity integration bounds.
    Consistency occurs with consumption of $< 3\%$ of the baryon fraction $\Omega_b$ after cosmic dawn.
	The results assume stellar first light at $z_\star=25$ \citep[][Fig 25]{harikanegoldrush2022} but are typical of the scenario for $15 < z_\star < 35$.
    Also indicated are the redshifts probed by $21$cm experiments.
}
\end{figure*}
We consider five high-redshift samples, and one local sample, of elliptical galaxies given by \citet{farrah22b}.
For the high-redshift samples we use: two from the WISE survey (one at $\widetilde{z} =0.75$ measured with the H$\beta$ line, and one at $\widetilde{z}=0.85$ measured with the \ion{Mg}{2} line), two from the SDSS (one at $\widetilde{z}=0.75$ and one at $\widetilde{z} =0.85$, with H$\beta$ and \ion{Mg}{2}, respectively) and one from the COSMOS field (at $\widetilde{z} = 1.6$). 
We then determine the value of $k$ needed to align each high redshift sample with the local sample in the $M_{BH} - M_{*}$ plane.  
If the growth in BH mass is due to cosmological coupling alone, regardless of sample redshift, the same value of $k$ will be recovered. 

To compute the posterior distributions in $k$ for each combination, we apply the pipeline developed by \citet{farrah22b}, which we briefly summarize.
Realizations of each galaxy sample are drawn from the sample with its reported uncertainties. 
The likelihood function applies the expected measurement and selection bias corrections to the realizations, as appropriate for each sample. The de-biased, and so best actual estimate, BH mass of each galaxy is then shifted to its mass at $z=0$ according to Equation~\ref{eqn:coupling} with some value of $k$.
Using the Epps-Singleton test, an entire high-redshift realization is then compared against a realization of the local ellipticals, where BH masses are shifted to $z=0$ in the same way.
The result is a probability that can be used to reject the hypothesis that the samples are drawn from the same distribution in the $M_{BH} - M_{*}$ plane, i.e. that they are cosmologically coupled at this $k$.

\section{Results \& Discussion}
We present posterior distributions in $k$, for each high-redshift to local comparison, in the top row of Figure \ref{fig:couplingpost}.
The redshift dependence of mass growth translates to the same value $k\sim 3$ across all five comparisons, as predicted by growth due to cosmological coupling alone. 
As further verification, we compute $k$ from a comparison between high-redshift WISE and COSMOS samples.
This comparison requires no BH bias corrections.
We find a consistent value of $k=2.96^{+1.65}_{-1.46}$.
Combining the results from each local comparison gives,
\begin{align}
  k=3.11^{+1.19}_{-1.33} \qquad \text{(90\% confidence)}\label{eqn:k},
\end{align}
which excludes $k=0$ at $99.98\%$ confidence, equivalent to $> 3.9\sigma$ observational exclusion of the singular Kerr interior solution.

\subsection{Implications}
Our result provides a single-channel explanation for the disparity in SMBH masses between local ellipticals and their 7-10~Gyr antecedents \citep{farrah22b}. 
Furthermore, the recovered value of $k \sim 3$ is consistent with SMBHs having vacuum energy interiors.
Our study thus makes the existence argument for a cosmologically realistic BH solution in GR with a non-singular vacuum energy interior.

Equation~(\ref{eqn:coupling}) implies that a population of $k \sim 3$ BHs will gain mass proportional to $a^3$.
Within a RW cosmology, however, all objects dilute in number density proportional to $a^{-3}$.
When accretion becomes subdominant to growth by cosmological coupling, this population of BHs will contribute in aggregate as a nearly cosmologically constant energy density.
From conservation of stress-energy, this is only possible if the BHs also contribute cosmological pressure equal to the negative of their energy density, making $k \sim 3$ BHs a cosmological dark energy species.

\subsection{$\Omega_{\Lambda}$ from the cosmic star formation history}
\label{sec:sfrd_summary}
If $k\sim3$ BHs contribute as a cosmological dark energy species, a natural question is whether they can contribute all of the observed $\Omega_{\Lambda}$.
We test this by assuming that: (1) BHs couple with $k=3$, consistent with our measured value; (2) BHs are the only source for $\Omega_{\Lambda}$, and (3) BHs are made solely from the deaths of massive stars.
Under these assumptions, the total BH mass from the cosmic history of star formation (and subsequent cosmological mass growth) should be consistent with $\Omega_{\Lambda} = 0.68$.

In Appendix~\ref{sec:sfrd} we construct a simple model of the cosmic star formation rate density (SFRD) that allows exploring combinations of stellar production rate, stellar IMF, and accretion history.
Figure \ref{fig:couplingsfr} displays models that produce the Planck measured value of $\Omega_\Lambda = 0.68$ \citep{aghanim2020planck} with the indicated IMF and Eddington ratio $\lambda_e$.
Any monotonically decreasing path inside the filled region produces $\Omega_\Lambda=0.68$ for some observationally viable IMF and accretion history, consuming at most $3\%$ of baryons.
This baryon consumption is compatible with the results of  \citet{macquartfrbbaryons2020}, who find that $\Omega_b$ at low redshift agrees with $\Omega_b$ inferred from the Big Bang to within $50\%$.

It follows from Equation~(\ref{eqn:coupling}) that cosmological coupling in BHs with $k = 3$ will produce a BH population with masses $>10^2M_\odot$.
If these BHs are distributed in galactic halos, they will form a population of MAssive Compact Halo Objects (MACHOs).
In Appendix~\ref{sec:machos}, we consider the consistency of SFRDs in Figure~\ref{fig:couplingsfr} with MACHO constraints from wide halo binaries, microlensing of objects in the Large Magellanic Cloud (LMC), and the existence of ultra-faint dwarfs (UFDs).
We conclude that non-singular $k=3$ BHs are in harmony with MACHO constraints while producing $\Omega_\Lambda = 0.68$, driving late-time accelerating expansion.

\section{Future Tests}
Further tests of cosmological coupling in BHs are essential to confirm or refute our proposal.
We list some examples below, highlighting possible tensions.

\subsection{Signatures in the cosmic microwave background}\label{sec:cmb}

A population of $k\sim 3$ BHs are a dark energy species.
Thus, their distribution in space need not trace baryons or dark matter at all times.
For example, \citet{croker2020implications} study the spatial distribution of $k \sim 3$ BHs with anisotropic stress \citep[e.g.][]{cattoen2005gravastars} at first order in cosmological perturbation theory.
Anisotropic stress within individual BHs leads to an effective fluid at first-order that resists clustering and can even drive the spatial distribution to uniformity.%
\footnote{
  For mathematical details, see \citet[][]{croker2022well}}
Resistance to clustering is computed from the non-singular BH model, of which there is currently none preferred (\S\ref{sec:exact_solutions}).
Some amount of anisotropic stress is, however, necessary to satisfy constraints on the galaxy two-point correlation function \citep{croker2020implications}.

Anisotropic stress from cosmologically coupled BHs at $z_\star = 25$ would maximally impact the CMB via the Integrated Sachs-Wolfe effect at $\ell = 5$, with lesser impacts at $\ell \gtrsim 5$ \citep[e.g.][]{koivisto2006dark,dodelsoncosmology2020}.
The low $\ell \lesssim 30$ of the CMB $TT$ anisotropy spectrum are anomalous in several respects \citep[e.g.][]{schwarzanomalies2016} and with amplitude in excess of cosmic variance at $\ell = 5$ \citep[e.g.][]{planck2018cmb}.
Any imprint of BH production at cosmic dawn on the CMB low $\ell$'s provides a precision test of non-singular BH models, and could enable independent constraint of the high-$z$ SFRD.

\subsection{Strong lensing of $\gamma$-ray bursts}
\label{sec:grbs}
The prevalence of strong gravitationally lensed GRBs has been used to estimate the comoving density of Intermediate Mass Black Holes (IMBHs) (\citealt{paynter21}, see also e.g. \citealt{wang21grav,yang21grav,chen22grav,lin22grav,kalan22,liao22rev}).
GRBs probe IMBHs at scales much greater than the non-linearity scale of $\sim 6~\mathrm{Mpc}$, complementing $\sim 50~\mathrm{kpc}$ MACHO constraints (Appendix~\ref{sec:machos}).
At these larger scales, the prediction of uniform spatial distribution from anisotropic stress (\S\ref{sec:cmb}) is directly applicable.

\citet{xiao2011redshift} report GRBs with spectroscopic redshifts and find that $90\%$ of GRBs have redshifts $z > 0.25$, with $50\%$ occurring at $z > 1.5$.
Across these distances, $k\sim 3$ cosmological coupling of BHs strongly impacts the comoving mass density of lenses along the line-of-sight, reducing it by factors of at least $2-16\times$.
Existing analyses do not incorporate this impact of cosmological coupling on the optical depth.
Thus, a direct comparison of measured $\Omega_\mathrm{IMBH}$ against the simulated BH populations in Appendix~\ref{sec:machos} is not yet possible.
A promising future test is to incorporate cosmological coupling into the optical depth calculation.
In general, a spectroscopic redshift is not available for candidate lensed GRBs, significantly weakening measurement of $\Omega_\mathrm{IMBH}$ \citep[e.g.][]{paynter21}.
Consistency with our population predictions, when evaluated at the lens redshift, could help to constrain such measurements.

\subsection{Stellar mass BH-BH merger rates}
The population properties of observable binary BH mergers probe cosmological coupling because their inspiral time can be a significant fraction of the age of the universe.
If $k > 0$, then the component masses of binary BH systems observed by gravitational-wave detectors are not representative of the birth masses.
This impacts the interpretation of merging populations of BHs \citep{2111.03634} and the stochastic background \citep{arzoumanian2020nanograv,christensen2019stochastic}.
Coupled mass growth further leads to accelerated orbital decay, though the rate of this decay may be model dependent \citep[c.f.][]{cro20,hadjidemetriou1963two}.
This aspect affects both the observed mass spectrum and rate of binary BH mergers.
For example, in the absence of cosmological coupling, stellar-mass binary BH systems with semi-major axis $\gtrsim 0.3~\mathrm{AU}$ can only merge within a Hubble time when highly eccentric. 
Conversely, with $k \gtrsim 0.2$, systems with initial semi-major axes $0.1 < R < 10^4~\mathrm{AU}$ can merge within less than a Hubble time.
This can lead to significant increases in merger rate.
Competing with this effect, however, is that mergers can happen so quickly after remnant formation that they merge at a redshift beyond the detection limit of current generation observatories.

\subsection{Direct measurement from orbital period decay}
Direct measurement of altered orbital decay rate in binary BH systems with proposed space-based observatories like LISA \citep{armano2018beyond}, DECIGO \citep{sato2017status}, and Taiji \citep{gong2021concepts} is likely not possible because these observatories can only track orbital decay in the final few months.
Decade-scale electromagnetic observations of a pulsar-BH orbit, however, would provide sufficient precision to measure cosmological coupling directly \citep[e.g.][]{cro19,weisberg2010binarypulsar}.

\subsection{Stellar mass BHs and stellar evolution}
Mass shifts consistent with cosmological coupling have been proposed to exist in the merging binary BH population, explaining both the observed BH mass spectrum and the existence of BHs in the pair-instability mass gap, though with a smaller coupling strength of $k\sim0.5$ \citep{cro21}.
A coupling of $k=3$ and adopting contemporary stellar population synthesis estimates can lead to an overabundance of BHs with masses $>120M_\odot$.
While uncertainties in binary BH formation channels \citep{Mandel_2022,zevin21}, massive binary star evolutionary physical processes \citep{Broekgaarden_2022}, nuclear reaction rates \citep{2020ApJ...902L..36F}, supernova core collapse physics \citep{Patton_2020}, and SFRD and metallicity evolution \citep{arxiv.2206.10622} leave population model flexibility, there are known young BHs within X-ray binaries with mass $\sim 20M_\odot$ \citep[e.g.][]{millerjones2021cygnus}. 
If this BH mass is typical of young stellar remnants at $z \lesssim 5$, then the distribution of remnant binary semi-major axes and eccentricities becomes constrained so as not to produce overly massive BH-BH mergers.
An important test for $k=3$ BHs is whether such constraints are plausible.

\subsection{Theoretical considerations}
\label{sec:exact_solutions}
As described in \S\ref{sec:intro}, there are known exact solutions with each of the following properties: strong spin, arbitrary RW asymptotics, dynamical mass, and interior vacuum energy equation of state.
Our result implies the existence, within GR, of an exact solution with all of these properties.
Currently, there is no known solution that possesses all four, though there are known solutions with various combinations of two \citep[e.g.][]{guariento2012realistic, dymnikova2016regular}.
Finding solutions that feature all four properties is an important theoretical step forward.

It is also interesting that a link between BHs and late-time accelerating expansion has been independently suggested within frameworks distinct from GR. 
\citet{afshordi2008gravitational} and \citet{prewei09} adopt a gravitational $\ae$ther framework in which an effective cosmological constant emerges from quantum gravity effects at BH horizons. 
Notably, this yields a reduction of the quantum field theory tuning required, from over one hundred decimal places to only two. 
Their scenario is however distinct from ours; it has a different theoretical basis and does not, to our knowledge, predict that BHs gain mass as the scale factor increases.

\subsection{Validation of preferential growth of SMBHs}
  Further validation of the measured preferential growth of SMBHs by \citet{farrah22b} is an essential test of our proposal.   
  This can be done by: deepening understanding of the relevant biases, improving morphological determinations of the high redshift samples, testing combinations of traditional assembly pathways in simulations, and improving the accuracy of SMBH and stellar mass measures.
An important future test is to improve the statistics and reliability of the high redshift sample.
Doing so requires assembling a sample of $\geqslant 10^{3}$ AGN in elliptical hosts with low SFRs and reddenings over $0.7<z<2.5$.
The sample should have reliable measures of host stellar mass and consistent measures of SMBH mass with a subset that includes multi-epoch reverberation mapped measures.
Such a study would enable narrower redshift intervals to be used at $z>0.7$, giving finer discrimination between cosmological coupling and other processes.

\subsection{Quasars at $z>6$ and the SMBH mass function}
The existence of SMBHs at $z\gtrsim6$ \citep[e.g.][]{trakh15} with masses $>10^{9}$M$_{\odot}$ is challenging to explain via accretion and direct collapse models \citep[e.g.][]{inayoshi2020seeds,volonteri2021origins}.
Cosmological coupling with $k = 3$, starting at $z = 25$, provides a mass increase of a factor of 51 by $z = 6$ (Equation~\ref{eqn:coupling}).
This would ease tensions between BH growth models and observations of $z>6$ quasars, but it has not been shown that cosmological coupling is required to do so.
BH masses must also increase between $z \sim 6$ and $z = 0$ by a factor of $343$.
Comparison of the BH mass function in quasars at $z \sim 6$ with the BH mass function at $z = 0$ must be compatible with this minimum increase.
Generalization of the \citet{soltan1982masses} argument to $k=3$ coupling is a first step, though comparison of the high-end mass cutoff is not sufficient, because SMBHs may cease to accrete luminously above $\sim 10^{11}M_\odot$ \citep[e.g.][]{king2015big, inayoshi2016there, carr2021constraints}.

\section{Conclusions}
Realistic astrophysical BH models must become cosmological at large distance from the BH.
Non-singular cosmological BH models can couple to the expansion of the universe, gaining mass proportional to the scale factor raised to some power $k$.
A recent study of supermassive BHs within elliptical galaxies across $\sim 7$~Gyr finds redshift-dependent $8-20\times$ preferential BH growth, relative to galaxy stellar mass.
We show that this growth excludes decoupled ($k=0$) BH models at $99.98\%$ confidence.
Our measured value of $k=3.11^{+1.19}_{-1.33}$ at $90\%$ confidence is consistent with vacuum energy interior BH models that have been studied for over half a century.
Cosmological conservation of stress-energy implies that $k=3$ BHs contribute as a dark energy species.
We show that $k=3$ stellar remnant BHs produce the measured value of $\Omega_\Lambda$ within a wide range of observationally viable cosmic star formation histories, stellar IMFs, and remnant accretion.
They remain consistent with constraints on halo compact objects and they naturally explain the ``coincidence problem,'' because dark energy domination can only occur after cosmic dawn.
Taken together, we propose that stellar remnant $k=3$ BHs are the astrophysical origin for the late-time accelerating expansion of the universe.

\begin{acknowledgments}
We thank the referees for very helpful reports.  
We thank M.~Valluri (U. Michigan) for suggesting the bias-free high redshift crosscheck.
We thank the David C. and Marzia C. Schainker Family for their financial support of required computations.
M.~Zevin is supported by NASA through the NASA Hubble Fellowship grant HST-HF2-51474.001-A awarded by the Space Telescope Science Institute, which is operated by the Association of Universities for Research in Astronomy, Inc., for NASA, under contract NAS5-26555.
G.~Tarl\'e acknowledges support through DoE Award DE-SC009193.
V.~ Faraoni is supported by the Natural Sciences \& Engineering Research Council of Canada (grant 2016-03803).
J.~Afonso acknowledges financial support from the Science and Technology Foundation (FCT, Portugal) through research grants PTDC/FIS-AST/29245/2017, UIDB/04434/2020 and UIDP/04434/2020.

The National Radio Astronomy Observatory is a facility of the National Science Foundation operated under cooperative agreement by Associated Universities, Inc.
\end{acknowledgments}

\begin{appendix}

\section{SFRD bounding model}
  \label{sec:sfrd}
Here we construct a simple model to establish whether plausible SFRDs lead to $\Omega_{\Lambda}\simeq0.7$, if massive stellar collapse produces $k=3$ BHs.
For the overall form of the SFRD, we adopt the model of \citet{madaufragos2017} at $z\leqslant4$.
Between $z=4$ and cosmic dawn at $z=25$ we adopt a power-law for the SFRD with exponent $\alpha < 0$, matched continuously to the $z\leqslant4$ SFRD.
In units of $M_\odot\cdot\mathrm{Mpc}^{-3}\cdot\mathrm{yr}^{-1}$,
\begin{align}
  \dot{\rho}_*(z, \alpha) :=
  \begin{cases}
    0.01 \frac{(1+z)^{2.6}}{1 + \left[(1+z)/3.2\right]^{6.2}} & z \leqslant 4 \\
    (z/4)^\alpha \dot{\rho}_*(\alpha) & 4 < z \leqslant 25 \\
    0 & z > 25
  \end{cases} \label{eqn:sfr_model}.
\end{align}
where $\dot{\rho}_*$ is the time rate of production of stellar mass.
The choice of varying the power law slope at $z>4$ is motivated by the disparities in observed $\dot{\rho}_*$ over $4<z<10$.
Estimates from infrared \citep{mrr18} and $\gamma$-ray bursts \citep{kist09} can be 1-2\,dex higher than estimates from UV-dropouts \citep{donnandropouts2022, harikanedropouts2022, bouwens2022}.
Varying $\alpha$ at $z>4$ allows us to encompass SFRDs consistent with all extant data.

To account for variation in the stellar IMF with redshift, we divide the SFRD into a Population III (Pop III) epoch at $z > 7$ (e.g. \citet{inoue2014} but c.f. \citet{liu2020did}) and a standard epoch, characterized by a Kroupa IMF at $z \leqslant 7$.
We consider scenarios where the Pop III epoch is characterized by either a Kroupa IMF or the top-heavy \citet{harikanedropouts2022} IMF $\propto M^{-2.35}$ for $50 \leqslant M \leqslant 500$, and zero elsewhere.
Finally, we use the redshift-dependent, mean-metallicity model of \citet{madaufragos2017}, truncated at $Z_\odot$, to enable use of standard metallicity-dependent zero-age main sequence (ZAMS) mass to remnant mass, delayed, core-collapse supernovae prescriptions by \citet{fryerbelczynski2012}.

For simplicity, we apply post-remnant formation accretion, with a fixed Eddington ratio $\lambda_e$, to all Pop III BHs over a duration $t_e$,
  \begin{align}
  M(t_i + t_e) \propto M(t_i)\exp\left(\lambda_et_e/\epsilon\right), \label{eqn:eddington}
\end{align}
brief enough that the impact on instantaneous rate from cosmological coupling of accreted mass can be neglected.
Because $t_e$, $\lambda_e$, and $\epsilon$ are degenerate in Equation~(\ref{eqn:eddington}), there is effectively only one parameter.
\citet{farrah22} use simulation studies of galaxy mergers to estimate $t_e = 44\pm 22~\mathrm{Myr}$ at $\lambda_e > 1$ from observations of 21 infrared-luminous galaxy mergers at $z<0.3$.
This value is consistent with $\sim 10~\mathrm{Myr}$ estimates by \citet{madau2014super} for high $z$ BH seeds, so we adopt $t_e = 20~\mathrm{Myr}$.
The uncertainty in $t_e$ is large, e.g. \citet{safarz20} find Pop III stellar BH accretion across timescales ${\sim 0.5~\mathrm{Myr}}$.
\citet{tortosa22} measure $\lambda_e \simeq 472$ for a Seyfert 1 galaxy, while simulations by \citet{inayoshihyper2016} find stable accretion onto isolated Pop III BHs at $\lambda_e \simeq 5000$.
As we apply this accretion to all Pop III BHs in our model, we consider models within a more conservative $\lambda_e \leqslant 30$.
Given the large range of plausible $t_e$ and $\lambda_e$, we fix the efficiency $\epsilon = 0.16$, roughly $30\%$ of the Kerr BH limit \citep[e.g.][]{bardeen1970kerr}.

To compute viable SFRDs, we use a Monte Carlo approach.
Because high-mass remnants formed at high redshift can easily dominate $\Omega_\Lambda$, care must be taken to sufficiently sample these tails of the distributions.
  We divide the domains for IMF and SFRD distributions into distinct windows such that, in these windows, the distributions are guaranteed to give Poisson errors $\leqslant 1\%$ in the lowest probability bin.
The total collection of draws is eventually re-weighted by relative areas under the windowed IMFs and Equation~\ref{eqn:sfr_model}.
In each window, we draw $10^6$ ZAMS stellar masses from the appropriate IMFs and $10^6$ redshift-dependent metallicities.
Given the birth masses and redshifts, we approximate a redshift of stellar death using $t_\mathrm{life} \sim 10^{10} (M/M_\odot)^{-\tau}~\mathrm{yr}$, where $\tau = 2.5$ \citep[e.g.][]{harwit2006astrophysical}.
We discard draws that live beyond $z=0$, and then determine collapsed remnant masses.
The ZAMS mass to remnant mass model we adopt assumes that all stars drawn are single stars.
Consistent with \citet{kalogera1996maximum}, we regard any remnants with mass $> 2.7M_\odot$ as BHs.
To convert the resulting population of BHs into a cosmological density, we regard this drawn population as residing within one co-moving $\mathrm{Mpc}^3$.
We first reweight all draws by the probability of having come from their respective windows.
We then convert from this reweighted mass to a predicted mass by rescaling $\dot{\rho}_*(z=0)$ by the total mass in long-lived stars, neutron stars, and initial BH remnant mass, as described by \citet{madaufragos2017}.
We compute $\dot{\rho}_*(z=0)$ by integrating Equation~\ref{eqn:sfr_model} in time, and scaling by an IMF-appropriate gas return fraction $1-R$ (e.g. $R=0.39$ given by \citet{madaufragos2017} for Kroupa, and $1$ for Harikane).
We then divide the summed predicted mass in BHs at $z=0$ by the critical density today $3H_0^2/8\pi G$ to get $\Omega_\Lambda$.

\section{Massive compact halo object constraints}
\label{sec:machos}
\begin{figure*}
  \includegraphics[width=0.49\linewidth]{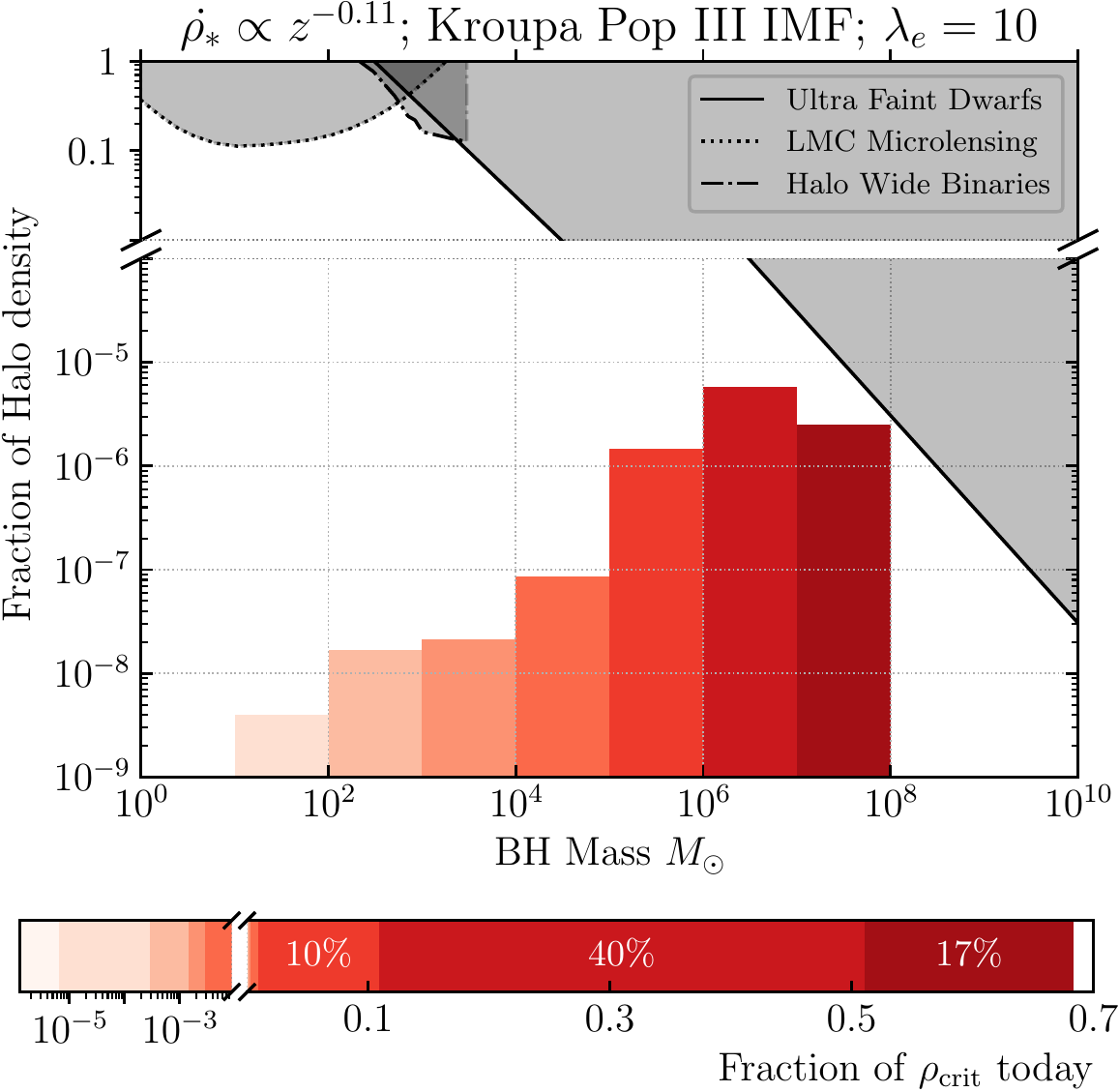}\hfill
  \includegraphics[width=0.49\linewidth]{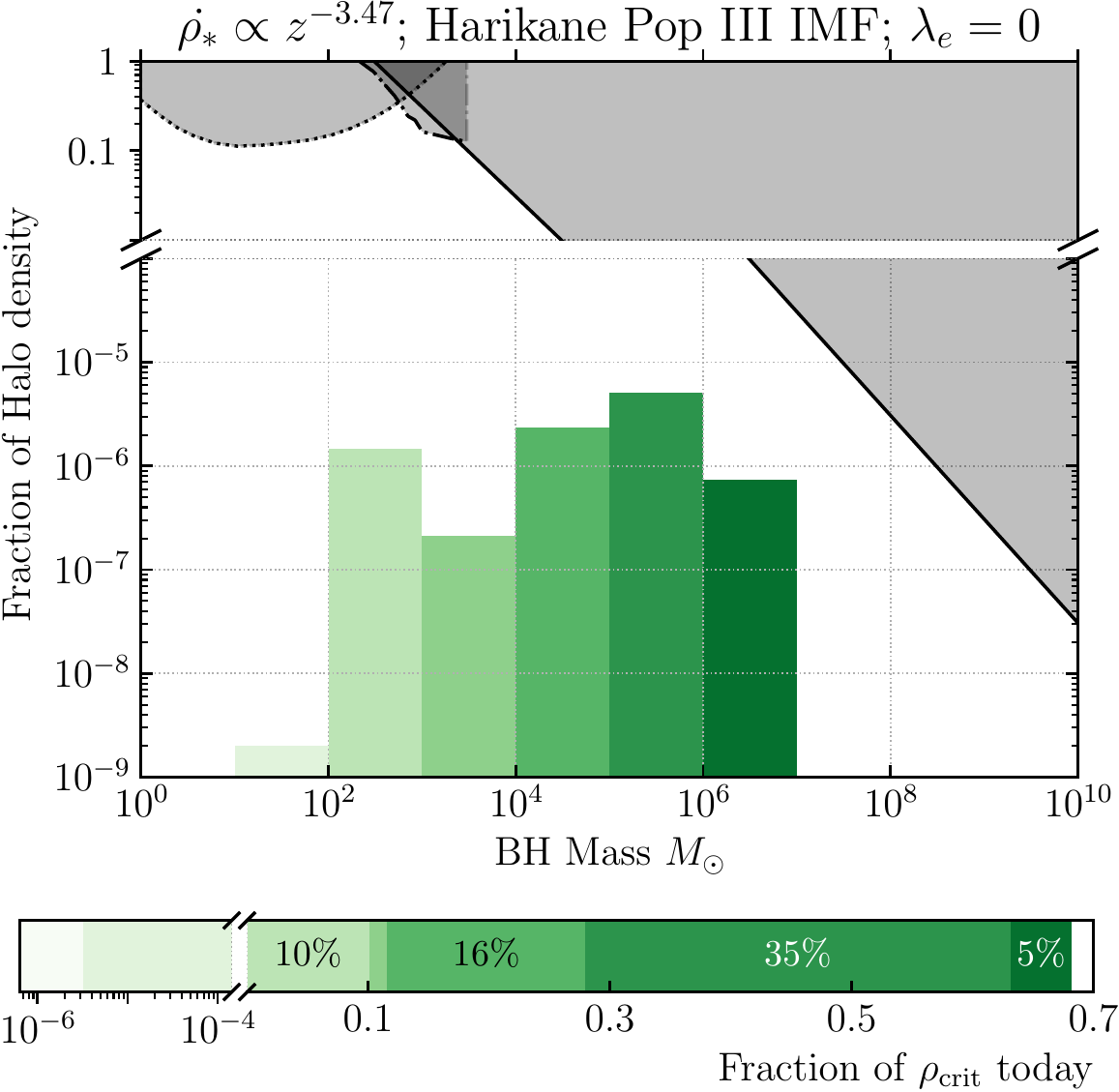} \\\par\bigskip
  \includegraphics[width=0.49\linewidth]{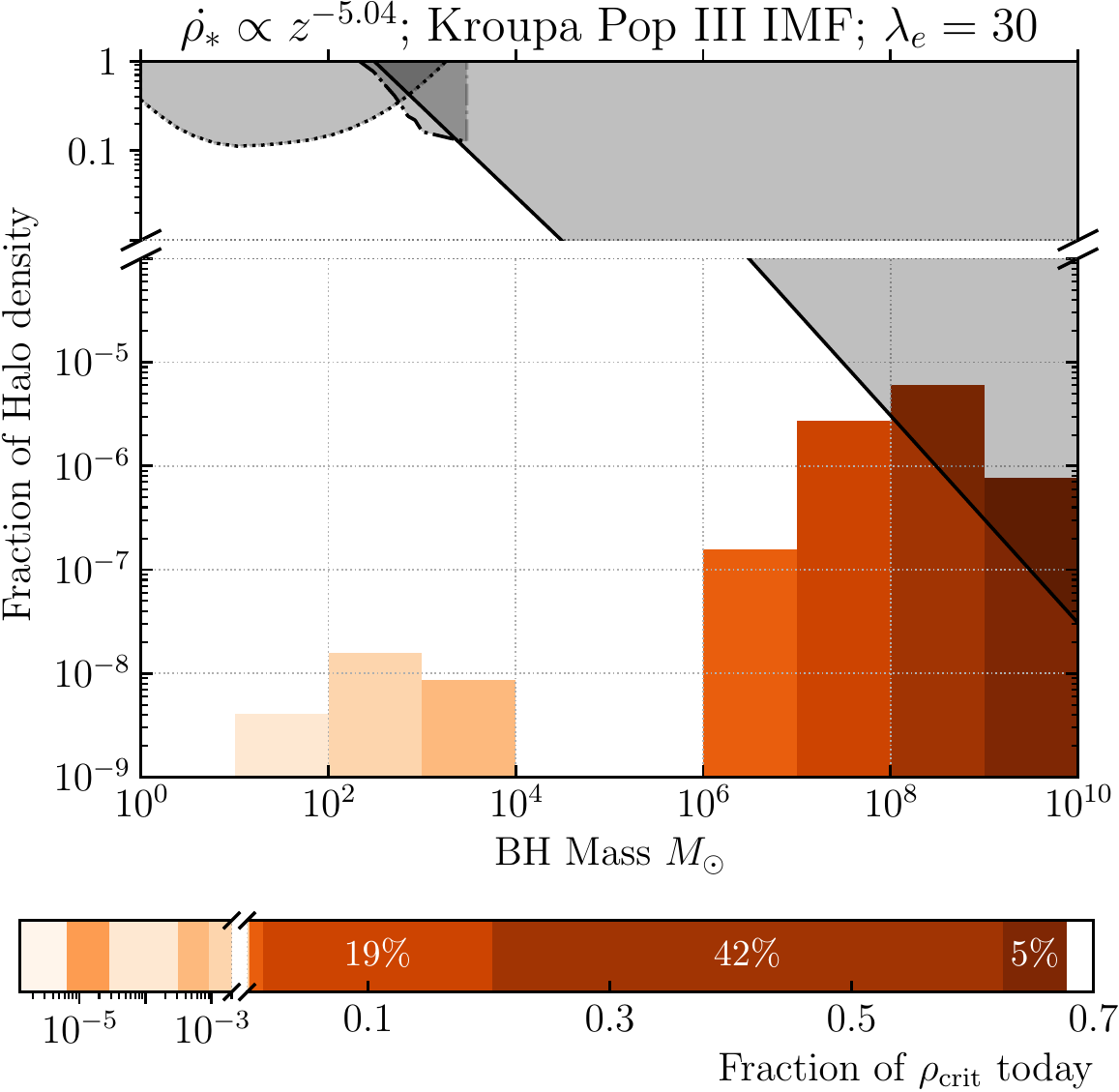}\hfill
  \includegraphics[width=0.49\linewidth]{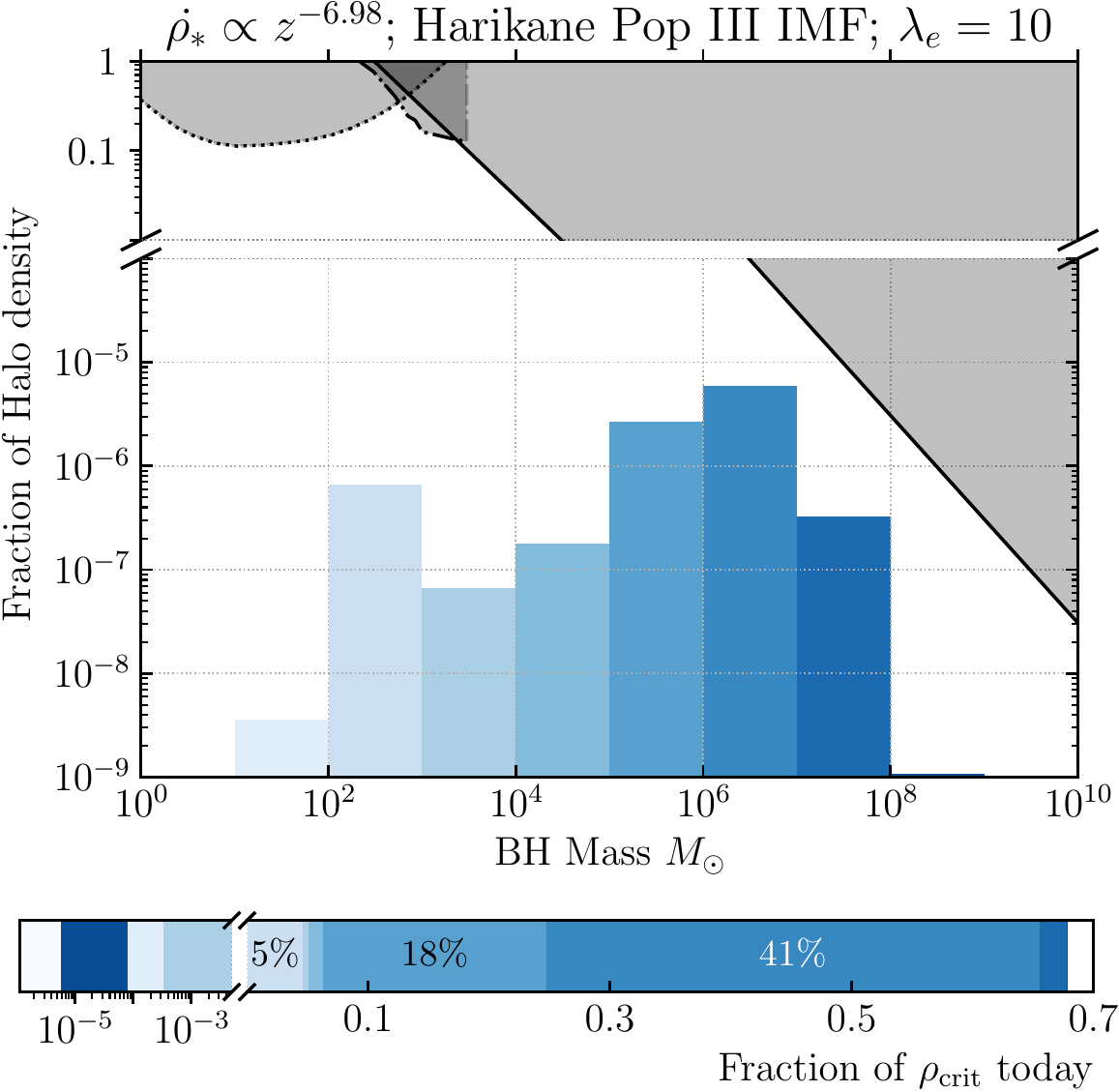}

\caption{\label{fig:couplingmacho}
  (Vertical bars) Fraction of $z=0$ halo density contributed by $k=3$ BHs, as produced by the indicated SFRDs in Figure \ref{fig:couplingsfr}.
    Models are ordered by increasing SFRD power-law slope, with colors set to agree with the model lines in Figure~\ref{fig:couplingsfr}.
    Also displayed (grey, shaded) are current constraints on MACHOs from microlensing \citep{blaineau2022mulensing}, wide halo binary disruption \citep{tylerwidebinary2022, monroyallen2014widehalo}, and ultra-faint dwarf (UFD) galaxy disruption \citep{brandt16}.
    Noting the broken vertical axis, the microlensing and halo binary constraints are easily satisfied.
    Dwarf galaxy constraints may discriminate candidate SFRD and IMF combinations.
    As shown, UFD constraints are overly conservative because they do not account for the decrease in comoving BH mass density at earlier times, as will be present in the $k>0$ coupled scenario.
    The effects of accretion at $z > 7$, as well as adopting a top-heavy IMF (Appendix~\ref{sec:sfrd}), are visible as decreased fraction in the IMBH range.
    (Horizontal bars) Fraction of present-day $\Omega_\Lambda$ contributed by each mass bin.
    Mass bins that contribute $< 5\%$ of $\Omega_\Lambda$ are unlabeled for clarity.
    Color gradients indicate mass binning and are common to both vertical and horizontal bars.
    Contributions less than $1\%$, including negligible contributions omitted in the vertical bars, are also shown in log scaling.
    Here, contributions are ordered by density fraction.}
\end{figure*}

Here we establish that a $k \sim 3$ BH population sufficient to produce $\Omega_\Lambda$ is consistent with constraints on massive compact halo objects (MACHOs).
In Figure \ref{fig:couplingmacho}, we display the contribution to halo density from $k=3$ BH masses at $z=0$ computed in the explicit SFRD models shown in Figure \ref{fig:couplingsfr}.
Consistent with the discussion in \S\ref{sec:cmb}, we have assumed a uniformly dispersed population in the computation of Figure~\ref{fig:couplingmacho}.
We adopt a uniform Milky Way (MW) halo density equal to the median DM density at the Solar System $8.8\pm0.5\times10^{-3}~M_\odot/\rm{pc}^3$ as measured by \citet{cautun2020milky}.
This is conservative relative to a mean halo density $1.7\times10^{-2}~M_\odot/\rm{pc}^3$, inferred from $1.37\times 10^{11}~M_\odot$ within the MW at $<20~\mathrm{kpc}$ as measured by \citet{posti2019mass}.

We display limits on MACHOs from microlensing \citep[e.g.][]{blaineau2022mulensing}, wide halo binary disruption \citep[e.g.][]{tylerwidebinary2022, monroyallen2014widehalo}, and UFD disruption \citep[e.g.][]{brandt16}.
We approximate the UFD constraint from \citet{brandt16} as $\propto 1/m$ following \citet[][Eqn.~(7.104)]{binney2011galactic} and convert from the UFD halo density assumed by \citet[][Figure~4]{brandt16} to our MW halo density scale by a multiplicative factor $0.3/0.0088$. 

All BH density contributions lie below the microlensing and wide binary limits by several orders of magnitude.
High accretion models with a Kroupa IMF show some tension with UFD constraints, which begin to have discriminatory power in our scenario.
Comparing the two top-heavy IMF models, the effect of accretion is to redistribute mass density from the IMBH range into the SMBH range.
The top-heavy IMF with no accretion model shows how a top-heavy IMF acts to populate the IMBH region.

The aforementioned constraints do not account for comoving BH mass density decrease with increasing redshift (Equation~(\ref{eqn:coupling})) for $k > 0$.
The constraints as presented are thus overly stringent because the probability of compact object interactions is reduced at earlier times.
The UFD galaxy and wide halo binary constraints are most impacted, as they consider the stability of astrophysical phenomena over Gyr timescales.
A thorough analysis that incorporates decreasing comoving density of compact objects further back in time is the subject of future work.

\end{appendix}

\bibliography{paper_couple}{}
\bibliographystyle{aasjournal}
\end{document}